\documentclass[10pt,twocolumn,letterpaper]{article}

\usepackage{cvpr}              % To produce the CAMERA-READY version
\usepackage{graphicx}
\usepackage{amsmath}
\usepackage{amssymb}
\usepackage{booktabs}
\usepackage{float}

\usepackage[pagebackref,breaklinks,colorlinks]{hyperref}

\usepackage[capitalize]{cleveref}
\crefname{section}{Sec.}{Secs.}
\Crefname{section}{Section}{Sections}
\Crefname{table}{Table}{Tables}
\crefname{table}{Tab.}{Tabs.}

\newsavebox\CBox
\def\textBF#1{\sbox\CBox{#1}\resizebox{\wd\CBox}{\ht\CBox}{\textbf{#1}}}

\def\cvprPaperID{9} % *** Enter the CVPR Paper ID here
\def\confName{CVPR}
\def\confYear{2024}

\title{Histopathological Image Classification with Cell Morphology Aware \\ Deep Neural Networks}

\author{Andrey Ignatov $^{1,2}$\\
{\tt\small andrey@vision.ee.ethz.ch}
\and
Josephine Yates $^2$\\
{\tt\small josephine.yates@inf.ethz.ch}
\and
Valentina Boeva $^2$\\
{\tt\small valentina.boeva@inf.ethz.ch}
\and \normalsize  ETH Zurich, Computer Vision Lab $^1$ \& Computational Cancer Genomics Group $^2$
}

\begin{document}
\maketitle
%%%%%%%%% ABSTRACT
\begin{abstract}

Histopathological images are widely used for the analysis of diseased (tumor) tissues and patient treatment selection. While the majority of microscopy image processing was previously done manually by pathologists, recent advances in computer vision allow for accurate recognition of lesion regions with deep learning-based solutions. Such models, however, usually require extensive annotated datasets for training, which is often not the case in the considered task, where the number of available patient data samples is very limited. To deal with this problem, we propose a novel DeepCMorph model pre-trained to learn cell morphology and identify a large number of different cancer types. The model consists of two modules: the first one performs cell nuclei segmentation and annotates each cell type, and is trained on a combination of 8 publicly available datasets to ensure its high generalizability and robustness. The second module combines the obtained segmentation map with the original microscopy image and is trained for the downstream task. We pre-trained this module on the Pan-Cancer TCGA dataset consisting of over 270K tissue patches extracted from 8736 diagnostic slides from 7175 patients. The proposed solution achieved a new state-of-the-art performance on the dataset under consideration, detecting 32 cancer types with over 82\% accuracy and outperforming all previously proposed solutions by more than 4\%. We demonstrate that the resulting pre-trained model can be easily fine-tuned on smaller microscopy datasets, yielding superior results compared to the current top solutions and models initialized with ImageNet weights. The codes and pre-trained models presented in this paper are available at: \url{https://github.com/aiff22/DeepCMorph}

\end{abstract}

%%%%%%%%% BODY TEXT
\section{Introduction}
\label{sec:intro}

\begin{figure*}[t!]
  \centering
  \vspace{-5mm}
   \includegraphics[width=0.99\linewidth]{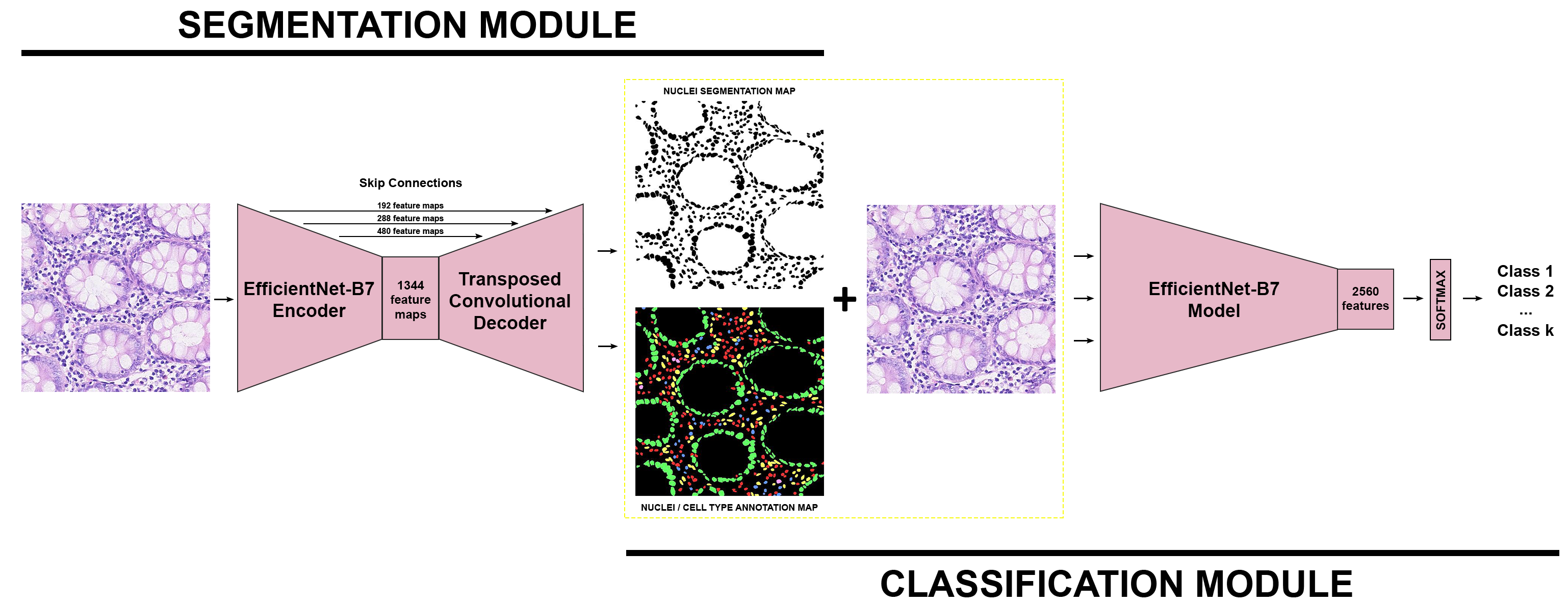}
   \caption{Overview of the proposed DeepCMorph network architecture. The model consists of two separate modules: the first one performs nuclei segmentation and cell type annotation. Its outputs are then stacked together with the original histopathology image and are passed to the second module performing the final classification task.}
   \vspace{-2mm}
   \label{fig:architecture}
\end{figure*}

Whole slide images (WSI), which are microscopy photos of stained tissue regions, are an important source of data for digital pathology. They contain rich visual information about tissue cell type composition, cell states, immune system activity, vasculature, cell abnormalities, \etc, and thus are widely used for the diagnosis and treatment selection for patients. Over the past years, histopathology image analysis has proven to be useful for such tasks as cancer screening and classification~\cite{komura2022universal,hou2016patch,tsaku2019texture,song2020clinically,yu2021large,bandi2018detection}, tumor grading~\cite{barbano2021unitopatho,lomenie2022can,bulten2022artificial,raju2020graph,koziarski2021diagset,xu2021computerized}, patient survival prediction~\cite{tsai2023histopathology,fu2020pan,yao2020whole,agarwal2021survival,shao2021weakly,wulczyn2020deep}, mitosis detection~\cite{nateghi2021deep,wang2014mitosis,balkenhol2019deep}, gene mutant prediction~\cite{qu2021genetic,coudray2018classification,yamashita2021deep}, tumor
immune infiltration quantification~\cite{xu2021deep,turkki2016antibody,abousamra2022deep}, gene expression~\cite{schmauch2020deep,dawood2021all} and biomarker prediction~\cite{wagner2023transformer,lu2022slidegraph+}, therefore making it crucial to develop performant WSI analysis tools.

While initially whole slide images were analyzed manually by pathologists, during the past years machine learning based approaches were gradually adopted for this task as they allowed for more accurate and efficient data processing. The first attempts were based on traditional computer vision methods~\cite{kather2016multi,shukla2017compact,valkonen2017metastasis} using handcrafted feature descriptors such as histograms of oriented gradients, Gabor filters, SIFT features, \etc. These approaches were later significantly outperformed by convolutional neural network (CNN) based models~\cite{banerji2022deep,van2021deep,srinidhi2021deep,wu2022recent,davri2022deep} that were trained in an end-to-end fashion and did not require any manual feature engineering. Recently, vision transformer (ViT) architectures were also proposed for WSI classification and retrieval problems, often demonstrating superior performance on these tasks~\cite{lazard2023giga,ikromjanov2022whole,xu2023vision,gul2022histopathological,zhang2022mc}. A typical approach to adapt these models for WSI analysis has been to use an existing neural network architecture and train it from scratch or initialize the model with weights obtained on the ImageNet dataset~\cite{komura2022universal,qu2021genetic,he2020integrating,weitz2022transcriptome,schmauch2020deep,kather2020pan}. Since numerous WSI datasets contain only tens or hundreds of samples, a significant limitation arises: while larger network architectures with millions of parameters may have the potential to learn more robust features and attain greater accuracy when trained on histopathology data, their practical performance is often constrained by insufficient training data and overfitting problems~\cite{kather2020pan}. One recently presented approach to deal with this problem is to pre-train models in an unsupervised manner using contrastive learning on a large corpus of diverse whole slide images. This led to a significant accuracy boost when tuning the obtained networks on WSI classification tasks~\cite{wang2022transformer,lu2023towards,wang2023retccl}. In this paper, we follow a different approach~--- rather than optimizing the network to generate representative features for each WSI patch, we suggest pre-training the model to learn cell morphology. Specifically, the model is trained to segment nuclei and classify cell types on histopathological slides. This information is directly integrated into the model and combined with the corresponding H\&E stained image to make predictions on the final downstream task.

Nuclei morphology plays an important role in understanding cell development and underlying cellular processes. Deviations from normal nucleus shape are often good indicators of external stress and thus are informative markers for a number of diseases, \eg breast cancers~\cite{chen2015new,nawaz2016computational}, carcinomas~\cite{fischer2020nuclear} or prostate cancers~\cite{carleton2018advances}. Nucleus morphology also changes during cell cycle progression and allows to identify actively dividing cells, which is a key characteristic of mutated cancer cells. Besides that, nuclei segmentation is additionally used for cell identification and counting: as accurate segmentation of the entire cell is very difficult in general due to invisibility of cell membranes (defining cell boundaries) on microscopy images, nuclei are usually used for recognition of single cells and their positions. Due to its vital role for histopathology image analysis, a large number of different nuclei segmentation datasets have been proposed over the past years~\cite{grossman2016toward,kumar2017dataset,sirinukunwattana2017gland,naylor2017nuclei,naylor2018segmentation,
vu2019methods,graham2019hover,graham2019mild,gamper2020pannuke,verma2021monusac2020,mahbod2021cryonuseg,verma2021monusac2020,graham2021lizard}. Various efficient deep learning-based solutions were also developed for this task~\cite{schmidt2018cell,graham2019hover,stringer10generalist,johnson2020automatic,hollandi2022nucleus,han2023ensemble,horst2024cellvit}, making it possible to achieve precise nuclei segmentation results for various cell types. In this work, we build the segmentation part of our model on top of the previous solutions, further enhancing its robustness by combining multiple datasets and applying extreme data augmentations.

\begin{figure*}[t!]
  \centering
  \vspace{-6mm}
   \includegraphics[width=0.84\linewidth]{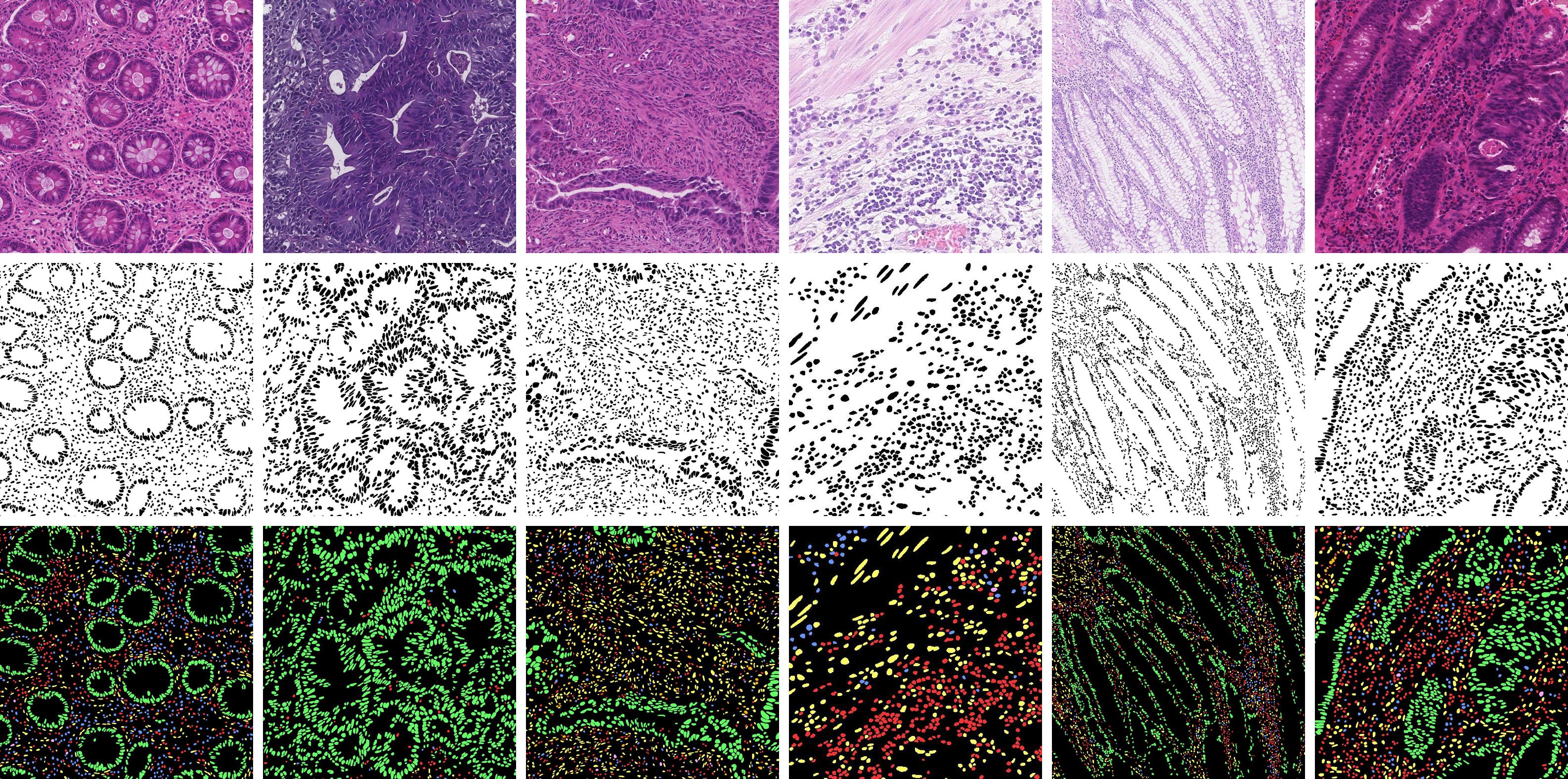}
   \caption{Sample images used for training the segmentation DeepCMorph module. Top row~-- original H\&E stained image patches, middle row~-- target nuclei segmentation maps, bottom row~-- cell annotation maps. For latter, red color encodes \textit{lymphocytes}, green: \textit{epithelial cells}, blue: \textit{plasma cells}, orange: \textit{neutrophils}, magenta: \textit{eosinophils}, yellow: \textit{connective tissue}.}
   \vspace{-4mm}
   \label{fig:dataset_sample}
\end{figure*}

Additionally, the cell type composition differs between healthy and diseased tissues and can serve to analyze disease progression. For instance, having large amounts of immune cells in a specific tissue region is generally associated with a strong inflammatory response~\cite{sherwood2004mechanisms}. In tumors, immune cell infiltration is strongly correlated with positive disease progression and improved patient survival~\cite{melssen2023barriers}. Therefore, an ability to distinguish between non-immune and immune cells and their subtypes might significantly improve the predictive power of the model. As several nuclei segmentation datasets provide additional cell type annotations~\cite{graham2019hover,verma2021monusac2020,graham2021lizard,gamper2020pannuke}, we explicitly use this information for training our segmentation module that generates both nuclei segmentation and cell type annotation maps. In this paper, we demonstrate that such information allows to achieve a significant accuracy boost, outperforming the latest transformer-based models trained on millions of histopathology images while also being more interpretable.

\smallskip
\smallskip

Our main contributions are:

\begin{itemize}
\setlength\itemsep{0.5mm}
\item We propose a novel histopathological image analysis DeepCMorph model that explicitly learns cell morphology: its segmentation module is trained to identify different cell types and nuclei morphological features.

\item The segmentation module is independent of the classification block; this allows to train it on all previously published nuclei classification and cell type annotation datasets and leverage this information when training the classification model part.

\item Unlike the recently proposed transformer-based foundation models~\cite{wang2022transformer,lu2023towards,wang2023retccl}, this solution is trained in a fully-supervised fashion, which lowers the training time and computational resources by an order of magnitude: for example, compared to the \textit{CTransPath}~\cite{wang2022transformer} model that requires 250 hours of training on 48 Nvidia V100 GPUs, our network can be trained on only one GPU in less than one week.

\item We apply extreme input data regularization to ensure that the model is not prone to learning any potential batch effect present in the data.

\item The model is pre-trained on 8 publicly available segmentation datasets and a large-scale Pan-Cancer TCGA dataset containing over 270K histopathological image patches. On this dataset, DeepCMorph established a new state-of-the-art result, bypassing all previous solutions by over 4\% of accuracy.

\item We show that that pre-trained model can be easily tuned on smaller datasets for tissue classification with excellent accuracy.

\item The proposed model has a fully convolutional architecture and can therefore be applied to images of arbitrary resolutions and aspect ratios.

\item We publicly release the codes and pre-trained models to facilitate the development of new performant histopathological image analysis tools.

\end{itemize}

%The remainder of the paper is structured as follows. In Section~\ref{sec:datasets} we describe all datasets used for training the model. Section~\ref{sec:methods} presents our architecture and training procedure. Section~\ref{sec:results} shows and analyzes the experimental results. Finally, Section~\ref{sec:conclusion} concludes the paper.

\section{Datasets}
\label{sec:datasets}

\begin{figure*}[t!]
\centering
\setlength{\tabcolsep}{1pt}
\resizebox{0.94\linewidth}{!}
{
\begin{tabular}{ccccccccc}
\includegraphics[width=0.25\linewidth]{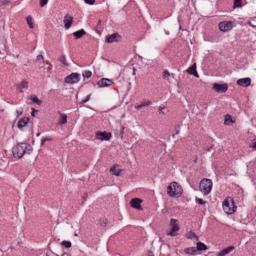}&
\includegraphics[width=0.25\linewidth]{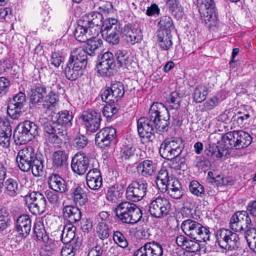}&
\includegraphics[width=0.25\linewidth]{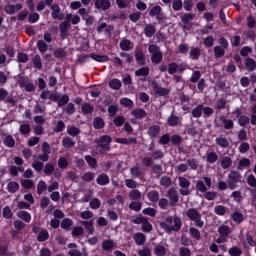}&
\includegraphics[width=0.25\linewidth]{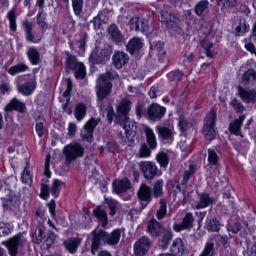}&
\includegraphics[width=0.25\linewidth]{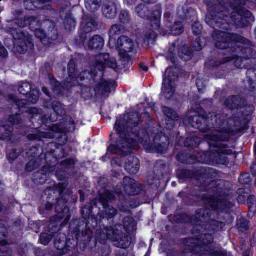}&
\includegraphics[width=0.25\linewidth]{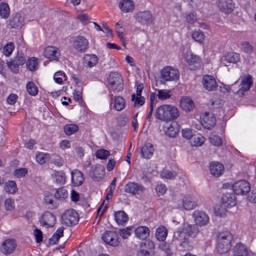}&
\includegraphics[width=0.25\linewidth]{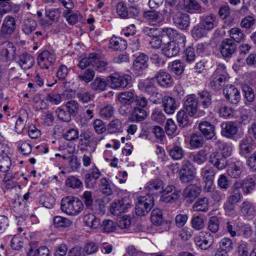}&
\includegraphics[width=0.25\linewidth]{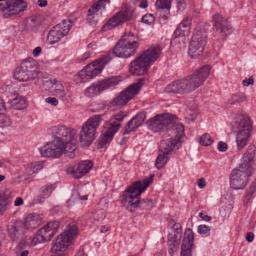}&
\\
Adrenocortical Carcinoma & Bladder Urothelial Carcinoma & Brain Lower Grade Glioma & Breast Invasive Carcinoma & Cervical Squamous Carcin. & Cholangiocarcinoma & Colon adenocarcinoma & Esophageal Carcinoma
\\
\includegraphics[width=0.25\linewidth]{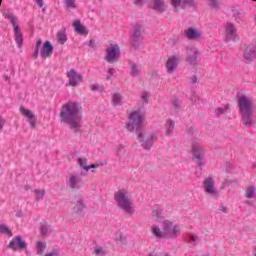}&
\includegraphics[width=0.25\linewidth]{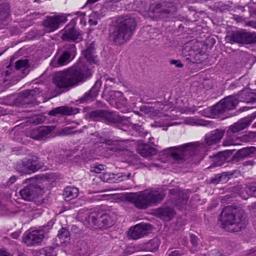}&
\includegraphics[width=0.25\linewidth]{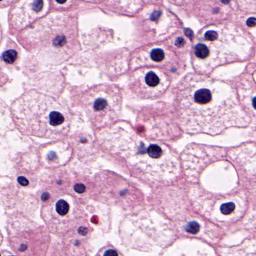}&
\includegraphics[width=0.25\linewidth]{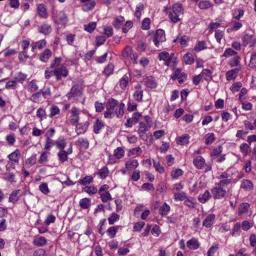}&
\includegraphics[width=0.25\linewidth]{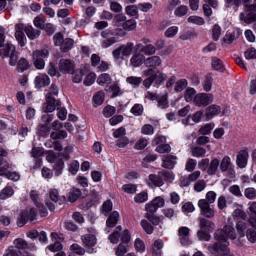}&
\includegraphics[width=0.25\linewidth]{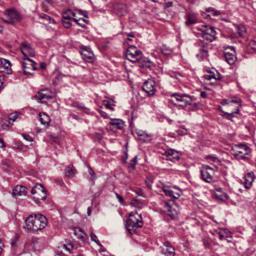}&
\includegraphics[width=0.25\linewidth]{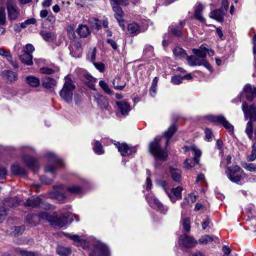}&
\includegraphics[width=0.25\linewidth]{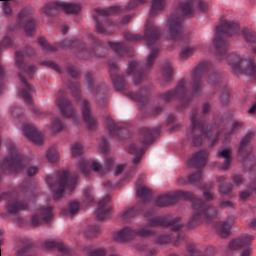}&
\\
Glioblastoma Multiforme & Squamous Cell Carcinoma & Kidney Chromophobe & Kidney Clear Cell Carcinoma & Kidney Papillary Carcinoma & Liver Hepatocellular Carcin. & Lung Adenocarcinoma & Lung Squamous Cell Carcin.
\\
\includegraphics[width=0.25\linewidth]{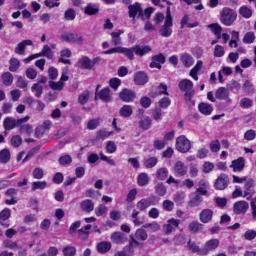}&
\includegraphics[width=0.25\linewidth]{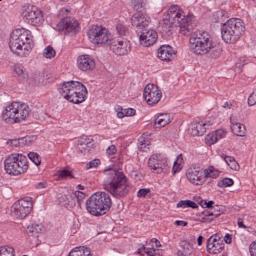}&
\includegraphics[width=0.25\linewidth]{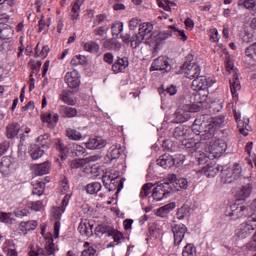}&
\includegraphics[width=0.25\linewidth]{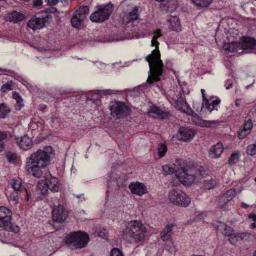}&
\includegraphics[width=0.25\linewidth]{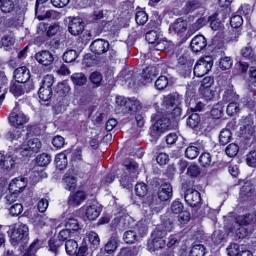}&
\includegraphics[width=0.25\linewidth]{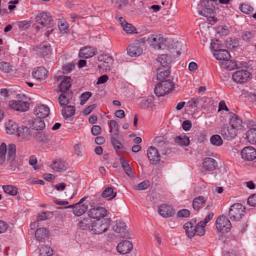}&
\includegraphics[width=0.25\linewidth]{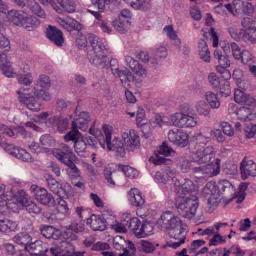}&
\includegraphics[width=0.25\linewidth]{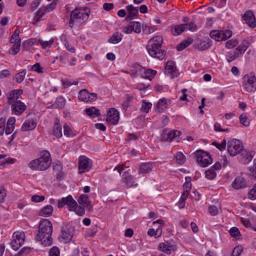}&
\\
Neoplasm B-cell Lymphoma & Mesothelioma & Ovarian Cystadenocarcinoma & Pancreatic Adenocarcinoma & Pheochromocytoma & Prostate Adenocarcin. & Rectum Adenocarcinoma & Sarcoma
\\
\includegraphics[width=0.25\linewidth]{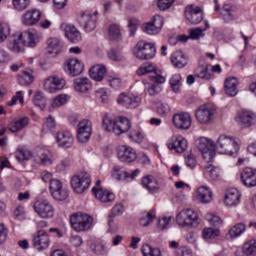}&
\includegraphics[width=0.25\linewidth]{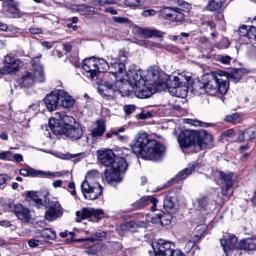}&
\includegraphics[width=0.25\linewidth]{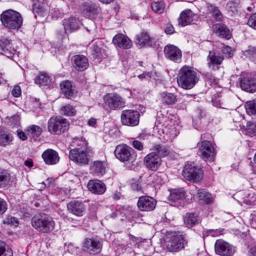}&
\includegraphics[width=0.25\linewidth]{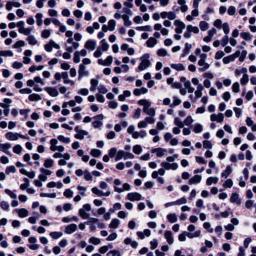}&
\includegraphics[width=0.25\linewidth]{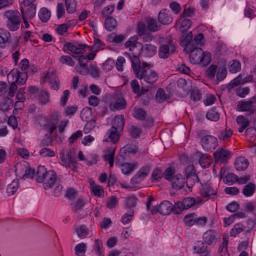}&
\includegraphics[width=0.25\linewidth]{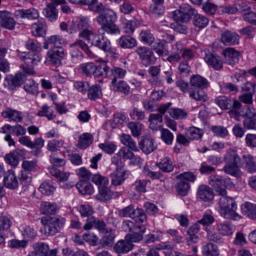}&
\includegraphics[width=0.25\linewidth]{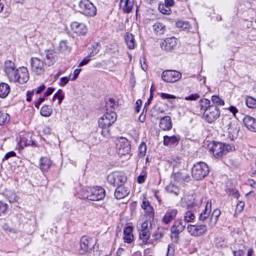}&
\includegraphics[width=0.25\linewidth]{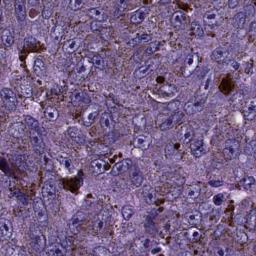}&
\\
Skin Cutaneous Melanoma & Stomach Adenocarcinoma & Testicular Germ Cell Tumors & Thymoma & Thyroid carcinoma & Uterine Carcinosarcoma & Uterine Endometrial Carcin. & Uveal Melanoma
\\
\end{tabular}
}
%\vspace{-2mm}
\caption{Sample H\&E stained image patches for 32 different cancer types from the Pan Cancer TCGA dataset~\cite{komura2022universal}.}
\vspace{-2mm}
\label{fig:TCGA_data}
\end{figure*}

This section describes nuclei segmentation, cell type annotation and tissue classification datasets that are used for training the segmentation and classification model blocks.

\subsection{Nuclei Segmentation and Annotation Datasets}

As our goal is to obtain a robust nuclei segmentation / cell type annotation model module that can be applied to tissues from various organs, we used the majority of previously published datasets proposed for these tasks. In particular, we combined data from 8 segmentation datasets described below to ensure a large diversity of tissue types, patients and imaging conditions:

\smallskip

\noindent $\bullet$ \textbf{Lizard dataset}~\cite{graham2021lizard} contains 291 H\&E stained histopathological images of colon tissue with 495179 labeled nuclei. Besides providing target nuclei segmentation maps, this dataset also annotates the type of the cell corresponding to each nucleus: \textit{epithelial cells, connective tissue cells, lymphocytes, plasma cells, neutrophils} and \textit{eosinophils}. The Lizard dataset is itself a combination of six different databases: \textBF{DigestPath}, \textBF{CRAG}~\cite{graham2019mild}, \textBF{GlaS}~\cite{sirinukunwattana2017gland}, \textBF{CoNSeP}~\cite{graham2019hover}, \textBF{PanNuke}~\cite{gamper2020pannuke} and \textBF{TCGA}~\cite{grossman2016toward} that were collected in total from 16 different centers and three countries, which ensures the diversity of the data from both biological and technical perspectives. The images are provided at 20$\times$ magnification.

\smallskip

\noindent $\bullet$ \textbf{CryoNuSeg dataset}~\cite{mahbod2021cryonuseg} contains 30 annotated H\&E stained images of resolution 512$\times$512 pixels with 7596 nuclei downloaded from the TCGA~\cite{grossman2016toward} database. Histopathological images from this dataset correspond to ten different human organs: \textit{adrenal gland, larynx, lymph node, mediastinum, pancreas, pleura, skin, testis, thymus} and \textit{thyroid gland}, and are provided at a magnification of 40$\times$.

\smallskip

\noindent $\bullet$ \textbf{MoNuSAC dataset}~\cite{verma2021monusac2020} contains 209 annotated H\&E stained images with 31411 nuclei downloaded from the TCGA~\cite{grossman2016toward} database. This data corresponds to 46 patients from 32 hospitals and four organs: \textit{breast, kidney, lung} and \textit{prostate}. The images are provided at 40$\times$ magnification.

\smallskip

\noindent $\bullet$ \textbf{BNS dataset}~\cite{naylor2017nuclei} contains 33 annotated H\&E stained images of resolution 512$\times$512 pixels with 2754 nuclei. The data corresponds to breast cancer, the images are provided at 40$\times$ magnification.

\smallskip

\noindent $\bullet$ \textbf{TNBC dataset}~\cite{naylor2018segmentation} contains 50 annotated H\&E stained images of resolution 512$\times$512 pixels with 4022 nuclei. The data corresponds to 11 patients and breast organs, the images are provided at 40$\times$ magnification.

\smallskip

\noindent $\bullet$ \textbf{KUMAR dataset}~\cite{kumar2017dataset} contains 30 annotated H\&E stained images of resolution 1000$\times$1000 pixels with 21623 nuclei. The data corresponds to 30 patients and seven organs: \textit{breast, kidney, liver, prostate, bladder, colon} and \textit{stomach}. The images are provided at 40$\times$ magnification.

\smallskip

\noindent $\bullet$ \textbf{MICCAI (CPM)--15/17 datasets}~\cite{vu2019methods} contain 79 annotated H\&E stained images with 10475 nuclei downloaded from the TCGA~\cite{grossman2016toward} database. The images correspond to four different cancer types: \textit{non-small cell lung cancer, head and neck squamous cell carcinoma, glioblastoma multiforme} and \textit{lower grade glioma}, and are provided at both 20$\times$ and 40$\times$ magnification.
\smallskip

\noindent $\bullet$ \textbf{PanNuke dataset}~\cite{gamper2020pannuke} contains 7901 annotated H\&E stained images of resolution 256$\times$256 pixels with 216400 nuclei from 19 different organs. Segmentation maps provided in this dataset were generated automatically and then revised by humans. Because of this, the targets for difficult or ambiguous cases are not very accurate~\cite{graham2021lizard} as they are limited by the performance of the used FCNN neural network. In this work, PanNuke dataset is used only for initial model pre-training.

\smallskip

Since the datasets use different approaches for storing images and annotations, they were first converted to the same data representation format: all data samples were transformed into 4-channel PNG images, where the first 3 channels encoded RGB image values, and the last one~-- the target segmentation and / or cell type annotation maps. This allowed to significantly simplify data pre-processing and reduce the size of the entire dataset from 40 GB to less than 2 GB. Overall, the combined dataset contains over eight thousand histopathological images, sample H\&E stained input data and target segmentation maps are shown in Fig.~\ref{fig:dataset_sample}. The instructions for downloading this data are provided on the official project webpage~\footnote{\url{https://github.com/aiff22/DeepCMorph}}.

\subsection{Tissue Classification Datasets}

\begin{figure*}[t!]
  \centering
  \vspace{-0.5cm}
   \includegraphics[width=0.99\linewidth]{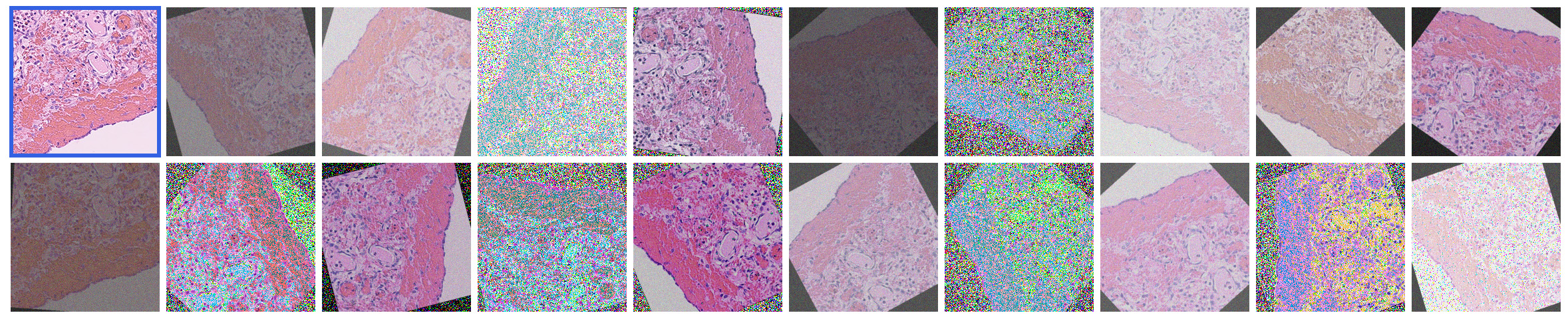}
   \vspace{-2mm}
   \caption{The original image patch (top left, denoted by blue frame) and training patches generated by the proposed data augmentations.}
   \vspace{-2mm}
   \label{fig:augmentations}
\end{figure*}

We consider four different histopathological datasets for training the classification DeepCMorph module. The large-scale Pan Cancer TCGA dataset is used for initial model pre-training and benchmarking its predictive capacity. The other three datasets are mainly used to assess DeepCMorph generalization abilities and its performance on other tissue subtypes. A detailed description of the considered datasets is provided below:

\smallskip

\noindent $\bullet$ \textbf{Pan Cancer TCGA dataset}~\cite{komura2022universal} was obtained by processing 8736 diagnostic slides downloaded from the TCGA~\cite{grossman2016toward} database and belonging to 7175 patients. For each WSI image, pathologists selected a number of representative patches corresponding to tumor regions. Overall, the dataset contains over 1.6 million patches extracted at 6 different magnification factors for 32 different cancer types, which list is provided in Fig.~\ref{fig:TCGA_data}. The authors performed color normalization using~\cite{reinhard2001color} to reduce the potential batch effect. Due to a large diversity, this dataset is perfectly suited for pre-training complex deep learning models for various histopathological tissue classification tasks. In this work, we consider the largest magnification factor of 20$\times$ (0.5 $\mu m$/pixel) and split the dataset randomly into training, validation and test parts using 70:15:15 ratio, which resulted in 188210, 41750 and 41750 patches, respectively. Notably, the splits were stratified per patient, \textit{i.e.}, images from a single patient were present only in the train or test set.

\smallskip

\noindent $\bullet$ \textbf{NCT-CRC-HE dataset}~\cite{kather2019predicting} contains 100K 224$\times$224 px image patches extracted from 136 colorectal adenocarcinoma samples from the National Center for Tumor Diseases (NCT). Nine tissue classes are present in the dataset: \textit{adipose, background, debris, lymphocyte, mucus, smooth muscle, normal colon mucosa, cancer-associated stroma} and \textit{colorectal adenocarcinoma epithelium}. All images are color normalized. A separate set of 7180 image patches from 50 patients with colorectal adenocarcinoma is used for testing.

\smallskip

\noindent $\bullet$ \textbf{Colorectal cancer (CRC) dataset}~\cite{kather2016multi} contains 5000 150$\times$150 px image patches corresponding to eight tissue types: \textit{epithelium, simple stroma, complex stroma, lymphoid follicles, debris, mucosal glands, adipose} and \textit{background}. Because of its small size, this dataset serves as a good benchmark for model generalization abilities. We split the dataset randomly into training, validation and test parts using 8:1:1 ratio.

\smallskip

\noindent $\bullet$ \textbf{UniToPatho dataset}~\cite{barbano2021unitopatho} contains 8699 patches of resolution 1812$\times$1812 px extracted from 292 WSIs. It was designed for colorectal polyp classification and adenomas grading task and has six tissue classes: \textit{normal tissue, hyperplastic polyp, tubular adenoma (low-grade and high-grade dysplasia)} and \textit{tubulo-villous adenoma (low-grade and high-grade dysplasia)}. Training / test data splits are explicitly provided by the authors.

\subsection{Data Augmentations}

Since the majority of our datasets contain heterogeneous data where samples corresponding to different classes (diseases) are often collected by different institutions, a pronounced \textit{batch effect} might be present in the data. Batch effect refers to the consequence of variations in tissue processing techniques across different labs or pathologists that may introduce distinct signatures specific to each site. These signatures can then be utilized to uniquely identify the corresponding WSIs.~\cite{howard2021impact}. Recently, Fang et al. demonstrated that neural networks can even recognize camera sensor models directly from images~\cite{fang2023sqad}, illustrating how microscopy equipment and image post-processing software might also introduce unique and identifiable signatures. Following the results from~\cite{fang2023sqad} indicating that such signatures are greatly destroyed by applying data augmentations, we adopted a similar strategy to mitigate any batch effect that might be present in the data. We used the following data augmentations:

\begin{itemize}
\setlength\itemsep{-0.02mm}
\item Random image rescaling (by 0--20\%),
\item Random change of image aspect ratio (by 0--10\%),
\item Random image rotation (by 0--360 degrees),
\item Random image sharpness adjustment (by 2$\times$),
\item Random image brightness adjustment (by 0--50\%),
\item Random image hue adjustment (by 0--10\%),
\item Random image contrast adjustment (by 0--70\%),
\item Random image saturation adjustment (by 0--30\%),
\item Addition of random Gaussian noise to images.
\end{itemize}

These augmentations are applied to all nuclei segmentation, cell type annotation and the Pan Cancer TCGA datasets. Sample results of such augmentations are demonstrated in Fig.~\ref{fig:augmentations}, indicating that color-, scale- and texture-related batch effect should be generally eliminated due to the considered extreme scales.

\section{Proposed Method}
\label{sec:methods}

This section provides the architectural and training details of the proposed DeepCMorph model (Fig.~\ref{fig:architecture}) that consists of two independent segmentation and classification modules described below.

\subsection{Segmentation Module}

We aimed to design a solution capable of processing arbitrary resolution images, thus we chose a fully convolutional segmentation model with a U-Net~\cite{ronneberger2015u} like structure. Its high-level architecture is shown in Fig.~\ref{fig:architecture}: a normal 3-channel RGB image is first processed by the EfficientNet-B7~\cite{tan2019efficientnet} encoder. Its outputs are then passed to a decoder module consisting of 4 upscaling blocks, each one with two subsequent convolution and one transposed convolution layers. Three skip connections are used to pass features obtained at the beginning of EfficientNet blocks 2, 3 and 4 with 192, 288, and 480 feature maps, respectively, to the decoder module. The decoder produces two outputs of the same  resolution as the input image: nuclei segmentation map (1 channel) and cell type annotation map (7 channels corresponding to 7 cell annotation classes). The model is trained with a combination of the Binary Cross Entropy (BCE) loss applied to nuclei segmentation maps and the Categorical Cross Entropy (CCE) loss function applied to cell type annotation maps:
\vspace{-1.2mm}
\begin{equation*}
BCE(\mathbf{y},\mathbf{p}) = -\frac{1}{N}\sum_i^N (y_i log(p_i) + (1-y_i)log(1-p_i)),
\end{equation*}
\vspace{-4mm}
\begin{equation*}
CCE(\mathbf{y},\mathbf{p}) = -\frac{1}{N}\sum_{i=1}^N\sum_{c=1}^C y_{i,c} \cdot log(p_{i,c}),
\end{equation*}
where $\mathbf{y}$ is a matrix of the ground truth labels and $\mathbf{p}$ is a matrix of predicted values for each pixel and sample.
The segmentation module was trained in multiple stages:
\vspace{-1mm}
\begin{enumerate}
\setlength\itemsep{-0.8mm}
\item First, the PanNuke dataset was used for initial model pre-training and then discarded during all later steps.
\item Next, the combined nuclei segmentation datasets were used for model training. At this stage, the cell type annotation data was ignored and only BCE loss used.
\item After convergence, the model was subjected to a further round of training with the addition of cell type annotation data. We used an equal amount of BCE and CCE loss at this step.
\item Pre-trained EfficientNet-B7 encoder model was further tuned on the Pan Cancer TCGA dataset~\cite{komura2022universal} independently of the decoder model to distinguish between 32 cancer types. Experimental results revealed that this procedure significantly improved the quality of the learned feature maps of the EfficientNet-B7 encoder.
\item Steps 2-3 were repeated again, this time using the EfficientNet-B7 encoder already trained on both nuclei segmentation and cancer tissue classification tasks.
\end{enumerate}
After segmentation module training, the module's weights were frozen and its predicted nuclei segmentation and cell type annotation maps were used as additional inputs to the classification DeepCMorph module.

\subsection{Classification Module}

The classification module of the DeepCMorph model is also based on the EfficientNet-B7 architecture. This model stacks the original RGB histopathology image with the segmentation and annotation maps from the segmentation module, thus its input has 11 feature maps (3 RGB + 1 nuclei segmentation + 7 cell type annotation). The outputs of the last EfficientNet-B7 global average pooling layer are passed to a dense layer with a softmax activation function that produces the final model predictions. We should note that this pooling layer allows the network to handle images of arbitrary sizes as it averages each of 2560 feature maps from the last model layer, thus producing 2560 features irrespective of the input image resolution. Dropout with a rate of 0.2 is additionally applied here to achieve higher feature robustness. The model is trained to minimize the Categorical Cross Entropy (CCE) loss function on classification tasks. The training process of this module also involved multiple stages, we provide a summary of all steps below:

\smallskip
\noindent \textBF{Step 1:} As random / ImageNet weights are not perfectly suited for model initialization for the considered tasks, we first pre-trained the classification module to perform the same nuclei segmentation and cell type annotation task as described in the previous section. We added an additional decoder block and trained the model using the same procedure (steps 1--5) as was applied to the segmentation module. The main difference here is that no skip connections between the encoder and decoder were used, given we targeted feature accumulation in the last (bottleneck) EfficientNet-B7 layer. At this stage, instead of the segmentation and annotation maps from the segmentation module, random noise was introduced. This allowed the EfficientNet-B7 model to learn to analyze the input histopathological rather than solely functioning as an autoencoder that compressed the input features and propagates them to the output layer.

\smallskip
\noindent \textBF{Step 2:} Once model pre-training on the segmentation task was done, random noise used in the previous step was replaced with real feature maps from the segmentation module, and the model was tuned for additional 5 epochs on the same segmentation task. The goal of this step is to let the model perform integration of the learned histopathological image processing features with the feature maps from the segmentation module, while not completely repurposing its filters to only encode the input segmentation / annotation maps. To determine the appropriate stopping point for training, we monitored the accuracy on the validation set. Following a small initial decrease, we noticed a rapid increase, indicating successful integration.

\smallskip
\noindent \textBF{Step 3:} During the last step, the decoder was removed from the classification module and the pre-trained EfficientNet-B7 model was finally tuned on the final downstream tissue classification task.

\subsection{Implementation Details}

The model is implemented in PyTorch~\cite{paszke2019pytorch} and trained on four \textit{Nvidia 2080 Ti} GPUs with 12 GB of RAM. At each step, model parameters were optimized using the Adam~\cite{kingma2014adam} algorithm with a learning rate of 1e--4 till model convergence (if not stated otherwise) and then additionally tuned with a learning rate of 2e--5. The segmentation module was trained on 224$\times$224 px image crops with a batch size of 36. The resolution of the input images received by the classification module depended on the task and dataset, for images of size 224$\times$224 px the batch size was set to 64. The entire DeepCMorph model has 89M parameters.

\section{Experimental Results}
\label{sec:results}

This section provides and analyzes experimental results obtained with the DeepCMorph model. First, we assess the performance of the segmentation module on nuclei segmentation / cell type annotation data. Next, we check the results obtained on the Pan Cancer TCGA dataset and compare DeepCMorph performance to the previously proposed solutions. Finally, we check the generalization ability of the proposed model on three other tissue classification datasets.

\subsection{Nuclei Segmentation and Annotation Results}

\begin{table}[t!]
\centering
\vspace{-2mm}
\resizebox{1.0\columnwidth}{!}
{
\begin{tabular}{l|ccc}
Method & Dice Score & Binary PQ & Multi PQ \\
\hline
\hline
DeepCMorph, Extreme Data Augmentations & 0.8365 & 0.714 & 0.340 \\
DeepCMorph, Moderate Data Augmentations & 0.8406 & 0.728 & 0.368 \\
\end{tabular}
}
\vspace{-2mm}
\caption{Nuclei segmentation (Dice Score) and annotation (Panoptic Quality Score) accuracy on combined segmentation datasets.}
\label{tab:segmentation_combined}
\vspace{-4mm}
\end{table}

\begin{table}[b!]
\centering
\vspace{-4mm}
\resizebox{0.8\columnwidth}{!}
{
\begin{tabular}{l|c}
Method & Dice Score \\
\hline
\hline
U-Net~\cite{graham2021lizard,ronneberger2015u} & 0.735 \\
Micro-Net~\cite{graham2021lizard,raza2019micro} & 0.786 \\
HoVer-Net~\cite{graham2021lizard,graham2019hover} & 0.828 \\
\hline
DeepCMorph [Segmentation Module] & \textBF{0.832} \\
\end{tabular}
}
\caption{Segmentation accuracy results on the Lizard~\cite{graham2021lizard} dataset.}
\vspace{-1mm}
\label{tab:lizard}
\end{table}

After the segmentation module was trained on a combination of the previously described datasets, we first briefly analyzed its performance to ensure a high quality of the generated segmentation and annotation maps. Numerical results obtained on the hold-out test set showing both the segmentation (Dice Score) and annotation (PQ Score) quality are presented in Table~\ref{tab:segmentation_combined}. They indicate that the model is able to accurately segment nuclei and annotate cell types for different human organs and tissue types, which was also confirmed by visual observations of segmentation results. One can also notice that the proposed excessive histopathological image augmentation approach leads to only minor accuracy degradations while making the model considerably more resistant towards potential batch effects.

We additionally benchmarked DeepCMorph segmentation performance separately on the Lizard~\cite{graham2021lizard} dataset. Numerical results shown in Table~\ref{tab:lizard} indicate that it performs comparably or slightly better than other conventional deep learning based segmentation methods. As the initial experiments demonstrated that the DeepCMorph tissue classification accuracy is almost unsusceptible to small nuclei segmentation and annotation errors as they can be tolerated by the classification module, this performance was considered as sufficient for our further experiments.

\subsection{Pan Cancer TCGA Data Classification}

\begin{table}[t!]
\centering
\resizebox{1.0\columnwidth}{!}
{
\begin{tabular}{l|cc}
Method & BA, \% & Accuracy,~\% \\
\hline
\hline
VGG16 based solution~\cite{komura2022universal} &  & 30.7 \\
ResNet-18 based solution~\cite{garciaclasificacion} &  & 33.5 \\
ResNet-18 based solution~\cite{alejandro2023intermediate} &  & 54.1 \\
CTransPath~\cite{wang2022transformer} features + SVM  &  & 73.38 \\
UNI (ViT-Large foundation model)~\cite{chen2024towards} & 65.7 & -- \\
EfficientNet-B7 (ImageNet initialized weights) &  & 75.89 \\
EfficientNet-B7 (Pre-trained on nuclei segmentation)\, &  & 78.72 \\
CTransPath~\cite{wang2022transformer} tuned on the TCGA dataset &  & 78.77 \\
\hline
DeepCMorph, Extreme Augmentations & 71.81 & 82.00 \\
DeepCMorph, Moderate Augmentations  & \textBF{72.79} & \textBF{82.73} \\
\end{tabular}
}
\vspace{-3mm}
\caption{Accuracy results on the Pan Cancer TCGA dataset~\cite{komura2022universal}. BA here stands for Balanced Accuracy score.}
\label{tab:TCGA}
\vspace{-5mm}
\end{table}

We first trained the model on the Pan Cancer TCGA dataset. As it contains 32 different cancer types and data from thousands of patients, it works as a severe benchmark for histopathological tissue classification models, being able to reveal the differences in their predictive capacity due to high task complexity. We compared the performance of our model against CNN-based solutions previously proposed for this dataset~\cite{komura2022universal,garciaclasificacion,alejandro2023intermediate}; a recent \textit{UNI}~\cite{chen2024towards} foundation vision transformer model pre-trained on 100K diagnostic H\&E-stained WSIs; \textit{CTransPath}~\cite{wang2022transformer} foundation transformer model pre-trained with contrastive learning on 15M unlabeled patches cropped from WSIs; two EfficientNet-B7 models: one with ImageNet initialized weights and one pre-trained on our combined segmentation datasets.

The results for all models are presented in Table~\ref{tab:TCGA}. DeepCMorph achieved an accuracy of 82.7\%, outperforming the second best CTransPath foundation model that was additionally fine-tuned on the considered dataset by almost 4\%. An advantage of 7\% was obtained over the UNI model when considering the balanced accuracy reported by the authors. The importance of learning cell morphology was demonstrated by EfficientNet-B7 results: the first model trained starting from ImageNet weights demonstrated an accuracy of only 75.9\%. Pre-training EfficientNet-B7 on nuclei segmentation and cell type annotation task improved the accuracy to 78.7\%, still leaving a gap of 4\% compared to the DeepCMorph that is using explicit annotations.

Same as with the DeepCMorph segmentation module, we can notice that the proposed extreme data augmentation leads to an accuracy drop of less than 1\%. The resulting model still outperforms all other solutions by over 3\% while being insensitive to variations in staining, noise, sharpness and lighting conditions, thus significantly reducing the impact of any potential batch effect present in the data.

\subsection{Results on the NCT-CRC-HE Dataset}

\begin{table}[t!]
\centering
\resizebox{1.0\columnwidth}{!}
{
\begin{tabular}{l|cc}
Method & BA, \% & Accuracy,~\% \\
\hline
\hline
DenseNet based solution~\cite{khvostikov2021tissue} & 90.3 & 92.9 \\
VGG19 based solution~\cite{kather2019predicting} & & 94.3 \\
Inception-v3 based solution~\cite{wang2021accurate} & & 94.8 \\
ResNet-50 based solution~\cite{sun2023automatic} & & 94.8 \\
VGG16 based solution~\cite{anju2023finetuned} & & 95.3 \\
CONCH (ViT-Base foundation transformer model)~\cite{lu2023towards}  & 93.0 & -- \\
iBOT (ViT-Large transformer model)~\cite{filiot2023scaling} & 94.4 & 95.8 \\
DINO (ViT transformer model)~\cite{kang2023benchmarking} & 94.5 & 95.9 \\
Ensemble of 4 models \scriptsize{(DenseNet, IncResNetV2, Xception and custom)}\normalsize~\cite{ghosh2021colorectal} &  & 96.16 \\
EfficientNet-B7 (ImageNet initialized weights) & 94.76 & 96.18 \\
Ensemble of 5 models \scriptsize{(Same as~\cite{ghosh2021colorectal} + VGG16)}\normalsize~\cite{kumar2023crccn} &  & 96.26 \\
CTransPath~\cite{wang2022transformer}  &  & 96.52 \\
\hline
DeepCMorph & \textBF{95.59} & \textBF{96.99} \\
\end{tabular}
}
\vspace{-2mm}
\caption{Accuracy results on the NCT-CRC-HE-7K validation dataset~\cite{kather2019predicting}. BA stands for Balanced Accuracy score.}
\label{tab:NCT_CRC}
\vspace{-5mm}
\end{table}

NCT-CRC-HE is a popular dataset with a large number of previously proposed deep learning solutions. We tuned the DeepCMorph model on this dataset, initializing it with weights obtained on the TCGA classification task, and assessed its performance on the conventional NCT-CRC-HE-7K test split. Table~\ref{tab:NCT_CRC} shows the results for different methods obtained on this task. With an accuracy of 96.99\%, DeepCMorph outperformed all other solutions, including visual transformer based CONCH~\cite{lu2023towards}, iBOT~\cite{filiot2023scaling}, DINO~\cite{kang2023benchmarking} and CTransPath~\cite{wang2022transformer} models that were pre-trained on a large cohort of histopathological data. We should additionally highlight a relatively good performance of the baseline ImageNet-initialized EfficientNet-B7 model that achieved an accuracy of 96.18\%, which is comparable to the results of more complex solutions. This also justifies the choice of the DeepCMorph backbone as the underlying EfficientNet-B7 model has a powerful architecture allowing to identify and learn complex patterns from histopathological data.

\subsection{Results on the NCT-CRC-HE Dataset}

\begin{table}[b!]
\centering
\vspace{-3mm}
\resizebox{1.0\columnwidth}{!}
{
\begin{tabular}{l|c}
Method & Accuracy,~\% \\
\hline
\hline
Conventional CV feature descriptors~\cite{kather2016multi} & 87.40 \\
Ensemble of 4 models \scriptsize{(DenseNet, IncResNetV2, Xception and custom)}\normalsize~\cite{ghosh2021colorectal} & 92.83 \\
VGG19 based solution~\cite{faust2018visualizing} & 93.58 \\
KimiaNet (DenseNet based model)~\cite{riasatian2021fine} & 96.80 \\
Ensemble of 6 models \scriptsize{(AlexNet, GoogleNet, VGG, ResNet, IncV3 and IncResV2)}\normalsize~\cite{nanni2021ensemble} & 97.60 \\
EfficientNet-B7 (ImageNet initialized weights) & 96.46 \\
%CTransPath~\cite{wang2022transformer} features + SVM  & 96.67 \\
CTransPath~\cite{wang2022transformer} & 98.20 \\
\hline
DeepCMorph & \textBF{98.33} \\
\end{tabular}
}
\vspace{-1mm}
\caption{Accuracy on the Colorectal Cancer (CRC) dataset~\cite{kather2016multi}.}
\label{tab:CRC8}
\vspace{-3mm}
\end{table}

DeepCMorph results on the Colorectal Cancer (CRC) dataset are shown in Table~\ref{tab:CRC8}. As this dataset contains only 625 image patches per class, we used it to check the generalization ability of the proposed model. Same as in the previous section, we initialized DeepCMorph with weights obtained on the TCGA classification task and tuned it for a few epochs on the CRC data. Despite the small number of training samples, the model was able to achieve top results on this task with an accuracy of 98.33\%. This shows that the filters and features learned by the model on the TCGA data are transferable to other histopathological tasks, making it a powerful tool for WSI data analysis and classification.

\subsection{Results on the UniToPatho Dataset}

\begin{table}[t!]
\centering
\resizebox{1.0\columnwidth}{!}
{
\begin{tabular}{l|cc}
Method & BA, \% & Accuracy,~\% \\
\hline
\hline
ResNet-18 based solution~\cite{barbano2021unitopatho} ($\sigma$ = 800) & 40.0 & -- \\
DeepCMorph (TCGA pre-trained) features + SVM \, & 42.51 & 46.31 \\
DeepCMorph tuned on UniToPatho & \textBF{47.35} & \textBF{55.81} \\
\end{tabular}
}
\vspace{-2mm}
\caption{Accuracy results on the UniToPatho dataset~\cite{barbano2021unitopatho}. Here, DeepCMorph is applied to WSI patches of resolution 1812$\times$1812 pixels. BA stands for Balanced Accuracy score.}
\label{tab:UniToPatho}
\vspace{-5mm}
\end{table}

UniToPatho dataset is used in different contexts including WSI retrieval~\cite{wang2023retccl} or hierarchical multi-scale WSI scan processing~\cite{deng2024cross}. In this work, we focus only on working with the provided 1812$\times$1812 px WSI patches to demonstrate that the DeepCMorph model can be used for high-resolution histopathological image data analysis, unlike transformer-based models restricted to a specific (usually small) input image size. Table~\ref{tab:UniToPatho} shows the results of the proposed solution obtained using two setups: when DeepCMorph model pre-trained on TCGA data is used to produce features for the considered patches that are later classified with SVM, and when it is additionally fine-tuned on smaller 256$\times$256 px crops and then applied to the original patches to provide direct predictions. Even the first setup managed to achieve a higher balanced accuracy (42.51\%) compared to the ResNet-18 model trained on UniToPatho in~\cite{barbano2021unitopatho}, which again demonstrates the versatility of the features learned by the DeepCMorph on TCGA data. After a short additional fine-tuning, the balanced accuracy improves further, reaching 47.35\%. Thus, DeepCMorph can be used to classify or annotate large-resolution patches or even entire WSI scans without a need for tile-based image processing and aggregation of the obtained crop-level results.

\section{Conclusion}
\label{sec:conclusion}

In this work, we considered the problem of histopathological image analysis and proposed a novel DeepCMorph model leveraging the understanding of cell morphology for more accurate tissue classification. DeepCMorph's segmentation module provides additional nuclei segmentation and cell type annotation predictions to the classification module, and is trained on a combination of 8 publicly available segmentation datasets. The classification module was first pre-trained on a large-scale TCGA dataset containing over 270K tissue patches for 32 different cancer types. The proposed DeepCMorph solution achieved the state-of-the-art results on four different tissue classification tasks, outperforming foundation transformer models like CTransPath pre-trained on millions of WSI patches. Due to a fully-convolutional architecture, the model can classify images of arbitrary sizes, avoiding the need for tile-base WSI processing. Finally, we open source the proposed solution and provide pre-trained models to facilitate the development of efficient histopathological image processing methods.

%%%%%%%%% REFERENCES
{\small
\bibliographystyle{ieee_fullname}
\bibliography{egbib}
}

\end{document}